\documentclass[bibyear]{aa}  

\usepackage{graphicx}
\usepackage{txfonts}
\usepackage{gensymb}
\usepackage{lipsum}
\usepackage{subcaption}         
\usepackage{lscape}             
\usepackage{placeins}           
\usepackage{natbib}
\usepackage{comment}
\usepackage{tablefootnote}
\usepackage[normalem]{ulem}
                                
\usepackage{hyperref}

\hypersetup{
    colorlinks=true, 
    citecolor=blue,  
    linkcolor=blue, 
    urlcolor=blue    
}

\def\medd{$\dot{m}_{\rm Edd}$}
\def\mbh{$M_{\rm BH}$}
\def\swift{{\it Swift}}

\def\spin{$\alpha^{*}$}
\def\ltransf{$L_{\rm transf}/L_{\rm disc}$}
\def\fcol{$f_{\rm col}$}

\begin{document}

   \title{X-ray disc reverberation modelling of the X-ray/UV/optical spectral/timing properties of Fairall 9}
   

   \author{M. Polioudakis\inst{1,2}\and
        \ M. Papoutsis\inst{1,3} \and
        \ I. Papadakis\inst{3,4}\and
        \ C. Panagiotou\inst{5} \and
        \ E. Kammoun\inst{6} \and
        \ M. Dovciak\inst{7}
        }

    \institute{Department of Physics, University of Crete, 71003 Heraklion, Greece
    \and Dipartimento di Fisica ‘Ettore Pancini’, Università Federico II, via Cinthia, I-80126 Napoli, Italy
    \and Institute of Astrophysics, FORTH, GR-71110 Heraklion, Greece 
    \and Department of Physics and Institute of Theoretical and Computational Physics, University of Crete, 71003 Heraklion, Greece
    \and MIT Kavli Institute for Astrophysics and Space Research, Massachusetts Institute of Technology, Cambridge, MA 02139, USA
    \and Cahill Center for Astronomy and Astrophysics, California Institute of Technology, 1200 California Boulevard, Pasadena, CA 91125, USA
    \and Astronomical Institute of the Academy of Sciences, Boční II 1401, CZ-14100 Prague, Czech Republic
    }

   \date{Received XXXX}

 
  \abstract
{Multiwavelength monitoring surveys of active galactic nuclei
(AGN) have revealed correlated variability observed in the X-ray, UV,
and optical bands. X-ray reverberation, arising from the absorption of X-rays 
illuminating the accretion disc, provides a self-consistent
physical framework for interpreting these observations and imposing
constraints on the geometry and energetics of accretion flows and X-ray
coronae.}
{We aim to apply the X-ray disc reverberation framework to the
Seyfert 1 galaxy Fairall 9, a well-studied  AGN with a clear line of sight to the accretion disc, to 
investigate whether this physical scenario can simultaneously account
for its observed spectral and timing properties, as probed by its mean
spectral energy distribution (SED), UV/optical power spectral densities
(PSDs), and interband time lags.}
{We used multiwavelength data from the 2018-2021 \textit{Swift} intensive
monitoring campaign to construct the mean X-ray/UV/optical SED and to
compute PSDs in all bands. We first modelled the broadband average SED
using \texttt{KYNSED}, which is a physical X-ray reverberation model assuming lamp-post geometry. The resulting best-fit parameter space was then used to
model the UV/optical PSDs and further constrain the physical parameters
of the system. Finally, we tested whether the observed interband time
lags are consistent with the model predictions for the parameter sets
that simultaneously reproduce both the SED and the PSDs.}
{X-ray illumination of the accretion disc can explain the broadband mean SED of Fairall 9. The UV/optical variations are likely driven by the variable X-rays that illuminate the disc, and not by short-timescale disc fluctuations of unknown physical origin. X-ray disc illumination and reverberation can explain the mean energy spectrum, the UV/optical power spectra, and the wavelength-dependent time lags simultaneously for a common set of physical parameters.}
{}


   \maketitle
   \nolinenumbers
%

\section{Introduction}

The standard model for the broadband spectral energy distribution (SED) of Active Galactic Nuclei (AGN) postulates the accretion of matter onto the super-massive black hole (SMBH) at the centre of the host galaxy, forming an optically thick and geometrically thin accretion disc \cite[][NT73 hereafter]{shakura73,NT73}. Viscous processes within the accretion disc remove angular momentum from the gas, enabling it to spiral inward towards the black hole (BH). As the gas accretes, the gravitational potential energy is converted into thermal energy, heating the disc material. The resulting high temperatures lead to thermal radiation, producing the strong ultraviolet (UV) and optical emission that characterises AGN.

AGN are also strong X-ray emitters. However, the temperature of a standard NT73 disc is not high enough to justify it as black-body emission from the inner disc in the case of SMBHs. It is generally believed that the X-rays are produced in a region close to the black hole, commonly referred to as the ``X-ray corona'', where high-energy electrons up-scatter via inverse Compton scattering the disc's photons into X-rays \citep[e.g.][]{Haardt93, Haardt94}. 
The exact geometry, location, and powering mechanisms of the X-ray corona are yet unknown and are the subject of open and active research.

A defining characteristic of AGN is the variability across a wide range of timescales, from hours to decades, observed in all wavelengths from X-ray to infrared \citep[see][for a recent review]{paolillo25}. The fastest and largest amplitude variations are observed in the X-ray band. If the corona illuminates and heats the disc, we would expect the disc to re-emit this absorbed energy in the UV/optical bands. Since X-rays are highly variable, we also expect the disc to vary accordingly, but with a delay corresponding to the light travel time of the X-ray photons as they reach the disc. This phenomenon is referred to as ``disc X-ray thermal reverberation'' \citep[e.g.][]{Kazanas01,cacket2007}. Many Seyferts have been observed over the last decade by \textit{Swift}, \textit{AstroSat} and \textit{HST} (to a lesser extent),  and ground-based telescopes (in some cases) simultaneously in X-rays and many UV/optical bands (see, e.g. \cite{Edelson19} and \cite{kumari23} for typical light curves that have resulted from the \swift\ and \textit{AstroSat} campaigns, respectively).
The main objective of these multi-wavelength campaigns is the study of the correlation between the X-ray/UV/optical variations in AGN. 

The target of one of the densest and longest monitoring campaigns was Fairall 9 (F9). 
This is a Seyfert 1 galaxy with a BH mass of $2\times10^8 M_{\odot}$ (\citealt{Bentz})\footnote{The AGN Black Hole Mass Database, \url{https://www.astro.gsu.edu/AGNmass/}}, at a distance of $D_L=209\ \mathrm{Mpc}$ ($z=0.046$; NASA/NED)\footnote{The NASA/IPAC Extragalactic database (NED), is funded by the National Aeronautics and Space Administration and operated by the California Institute of Technology, \url{https://ned.ipac.caltech.edu}}. F9 was observed almost daily by \textit{Swift} from May 2018 to February 2020 (i.e., for a period of $\sim 1.8$ years). {\it Swift} continued monitoring F9 for 11 more months, but at a lower cadence (once every $\sim 4$ days).
\cite{Hernandez} were the first to study the X-ray/UV/optical variability of the source using light curves from the first  9 months of the monitoring campaign (May 2018 - February 2019). Subsequently, \cite{edelson} released the reduced dataset of the entire campaign, which included light curves in eight bands, from X-rays to optical wavelengths. This is one of the best sampled and longest \textit{Swift} multi-wavelength monitoring campaigns of an AGN until now. 

\cite{partington24} studied {\it NICER} spectra of the source,  which were taken at a two-day cadence simultaneously with the \swift\ monitoring campaign. \cite{edelson} studied in detail the correlation between X-rays and UV/optical bands in Fairall 9 using the well-sampled 1.8-year-long light curves. They found that variations within the UV/optical spectral bands are well correlated, but with a delay. When time lags are measured with respect to the W2 filter, they increase with increasing wavelength, as predicted in the case of thermal X-ray reverberation. In fact, \cite{K23rev} fitted the time lags of F9 as measured by \cite{Hernandez} using the \cite{kammoun21a} X-ray reverberation model. The same time lags were also fitted by \cite{Netzer22}, assuming that the observed time lags are caused by the response of diffuse emission from radiation pressure supported clouds in the Broad Line Region (BLR) to the continuum. However, \cite{edelson} showed that this model cannot explain the observed time lags as computed using light curves from the 1.8-year  campaign \citep[see Fig. 6 in][]{edelson}.  

\cite{hagen24} fitted the mean SED of F9 from the first year of the monitoring campaign assuming a warm corona on top of the accretion disc and a hot inner region where the X-rays are emitted. 
They also assumed intrinsic disc fluctuations that vary on long time-scales and propagate into the hot inner corona, which generates fast variability. In this way, the slow variations
from the disc modulate the fast variations from the corona, and as a result, the light curves contain 
both long- and short-term trends. However, in order to explain the observed time lags \citep[as reported by][]{Hernandez} they also assumed UV (and X-ray) reverberation from a wind. They proposed that the increase of the time lag with increasing wavelength is due to the increased contribution of the (constant lag) bound-free continuum to the spectrum, rather than indicating intrinsically larger reverberation distances for longer wavelengths. In a recent work, \cite{hagen25} studied the SED variations, as well as the power spectrum and phase-lag spectrum variations of F9, using the full $\sim 3$ year long light curves. Their findings suggest that the inner structure of the accretion flow in this source may evolve on the timescales probed by the campaign.

We used data from the complete monitoring campaign of F9 to compute the mean SED and the well-sampled, 1.8-year light curves to estimate the power spectrum (PSD) in the X-ray and UV/optical bands.  Our first objective is to fit the mean SED using \texttt{KYNSED} \citep{dovciak22}. This is a theoretical model which is appropriate in the case of X-ray illuminated accretion discs. 
Then, following \cite{Panagiotou}, \cite{papoutsis25} and \cite{K23rev}, we investigate whether X-ray disc reverberation can also explain the observed power spectra (in the UV and optical bands) and the observed time lags, assuming the best-fit model parameters from the modelling of the mean SED.

The paper is organised as follows. In Section \ref{sect:data} we describe the data. In Section \ref{sect:mean_sed} we present the modelling of the mean spectral energy distribution. In Section \ref{sect:psdest} we outline the estimation of the observed UV/optical and X-ray power spectra. The X-ray reverberation model is introduced in Section \ref{sect: Xray_rev}, and its application to the power spectra is presented in Section \ref{subsec:fitting_PSDs}. In Section \ref{sect: timelags} we present the results from the model fitting of the observed time-lags, and in Section \ref{sect: discussion} we summarise and discuss our main results.

\begin{figure*}[ht]
\sidecaption
\includegraphics[width=12cm]{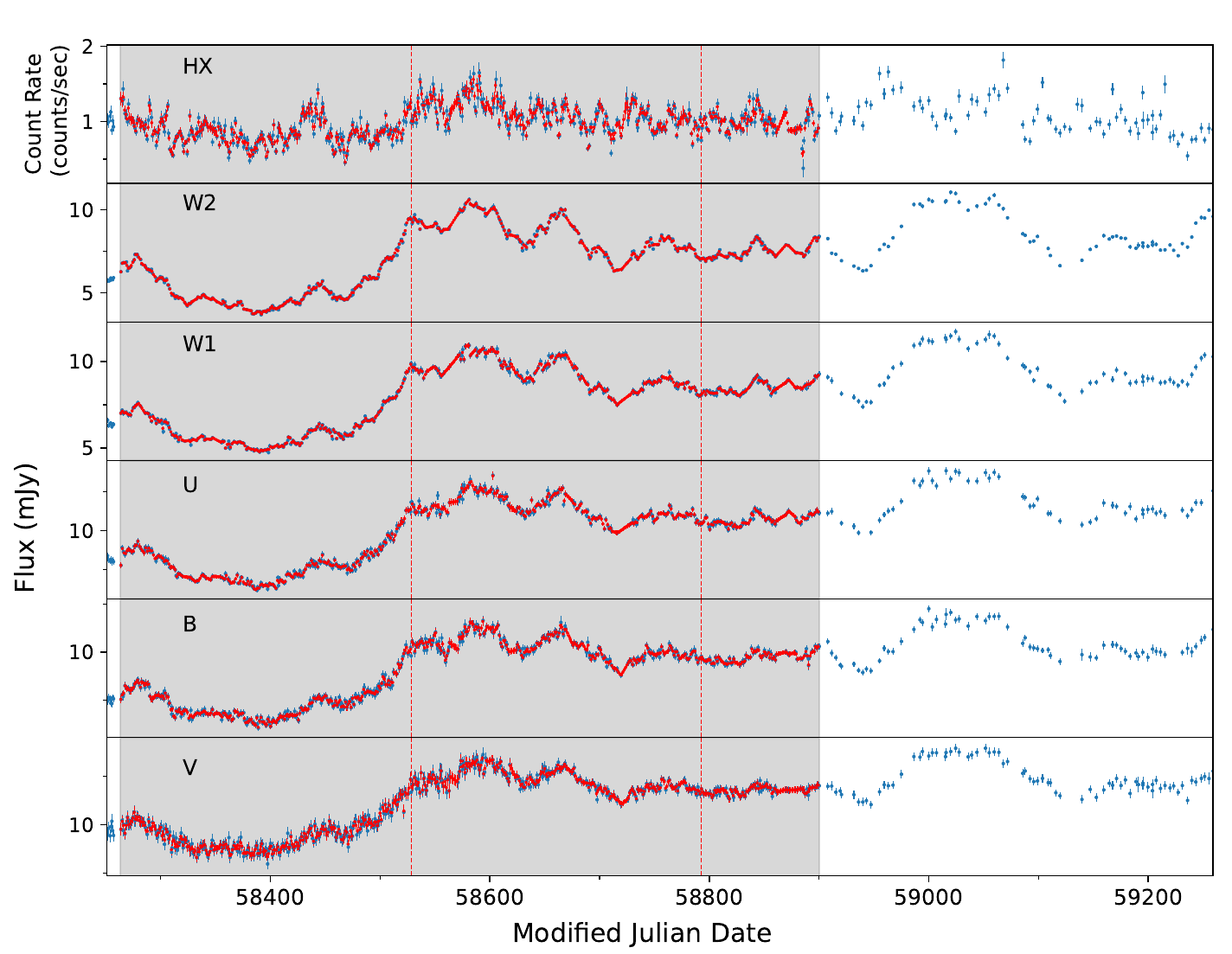}
  \caption{The \textit{Swift} light curves of Fairall 9 \citep[data are taken from][]{edelson}. Blue and red points indicate the observed and the interpolated light curves, respectively. The shaded area shows the part of the light curves we consider in the calculation of the PSD. Note that due to an early seven-day observational gap, the initial twelve days of observations were excluded from the analysis. The vertical dashed lines mark the two light curve parts we used to perform the stationarity test (Sect.\ref{subsec:stationarity}).}
     \label{fig:lightcurves}
\end{figure*}


\section{The data}
\label{sect:data}
We used X-ray and UV/optical data from \citet[][as listed in their Table\,2]{edelson}. The blue squares in Fig.\,\ref{fig:lightcurves} show the observed light curves in the X-rays (top panel) and in the UV and optical bands, as observed by \swift. Observations during the first 1.8 years (58251--58900 MJD) were acquired nearly daily, whereas the mean sampling interval in the latter period (58901--59259 MJD) was approximately four times larger. 

As we explain in Sect.\,\ref{sect:mean_sed}, we used data from the entire campaign to compute the mean SED (from X-rays to the optical bands). However, due to the sparser sampling of the light curves in the last part of the campaign, we considered the light curve parts within the shaded area in Fig.\,\ref{fig:lightcurves} for the calculation of the power spectrum. 
The mean time separation between successive observations, $\overline{\Delta t}_{\rm obs}$, and the number of points, $N$, in the 1.8 year-long light curves are listed in the third and fourth columns of Table \ref{table:lcinfo}. The second column lists the centroid wavelength and width of the UVOT filters and of the X-ray band (in keV), taken from \citealt{edelson}. We did not consider the M2 filter, because its transmission curve overlaps significantly with the W2 and W1 filters.

As we explain in Sect.\,\ref{sect:psdest}, we use the periodogram as an estimator of the power spectrum, which requires evenly spaced light curves. The spacing of the observations is not even, but, since the F9 light curves are regularly sampled, it is relatively straightforward to produce evenly sampled light curves using linear interpolation. The red circles in Fig.\,\ref{fig:lightcurves} indicate the interpolated light curves, which appear to be almost identical to the observed ones. Their errors were computed using the standard error propagation formula. The bin size of the interpolated light curves, $\Delta t_{\mathrm{int}}$, is equal to $\overline{\Delta t}_{obs}$, therefore, the number of points is the same in the observed and interpolated light curves. The median of the difference between the time of the observations and the time points in the interpolated light curves is listed in the last column of Table\ref{table:lcinfo}. It is small ($\sim$ 4 times smaller than the light curve bin size in all bands), and significantly smaller than the shortest timescale on which we estimate the power spectrum. 

\begin{table}[h!]
\caption{Summary of the \textit{Swift} observations we use for the calculation of the power spectra (i.e. the data points within the vertical dashed lines in Fig.\,\ref{fig:lightcurves}). }            
\label{table:lcinfo}     
\centering
\begin{tabular}{c c c c c}        
\hline\hline                 
Band  & $\lambda_0$ ($\Delta\lambda$)   & $\overline{\Delta t}_{\mathrm{obs}}$ & N   & $\Delta  t_{\mathrm{diff,med}}$ \\
 & (\AA) & (days) &  & (days) \\

\hline
   HX & 1.5-10 keV  & 1.12 & 567 & 0.27  \\
   W2 & 2055 (665)  & 1.35 &  472 & 0.32   \\
   W1 & 2580 (932)  & 1.46 &  434 & 0.37 \\ 
   U & 3463 (547)  & 1.41 &  451 & 0.34  \\ 
   B & 4350 (674)  & 1.26 &  504 & 0.34 \\ 
   V & 5425 (532)  & 1.21 &  526 & 0.29 \\ 
\hline                                  
\end{tabular}
\end{table}

\section{The average broadband SED}
\label{sect:mean_sed}

We used data from the total light curves (i.e., MJD 58251-59295) and computed the average flux and its error in the W2, W1, U, B, and V bands\footnote{We computed the error of the mean as $\sqrt{\sigma^2/N}$, where $\sigma^2$ is the variance of the light curve and $N$ is the number of points.}. We added a systematic error of 2\% of the mean flux, which accounts for the fact that the broad-band flux measurements may be affected by the contribution of strong emission lines, which are also variable, but not in sync with the continuum variations \citep[see Fig.\, 1 in][]{edelson}. For the X-ray part of the SED, we selected \textit{Swift}/XRT observations 
whose count rate falls within one standard deviation of the mean count rate (there are 30 such observations). Using these observations, we extracted an average X-ray spectrum with the \textit{Swift}/XRT automatic data product generator\footnote{\url{https://www.Swift.ac.uk/user_objects/}} \citep{evans}. The spectrum was grouped to 50 counts per bin. 

\begin{figure}[ht]
    \centering
    \includegraphics[width=1.1\linewidth]{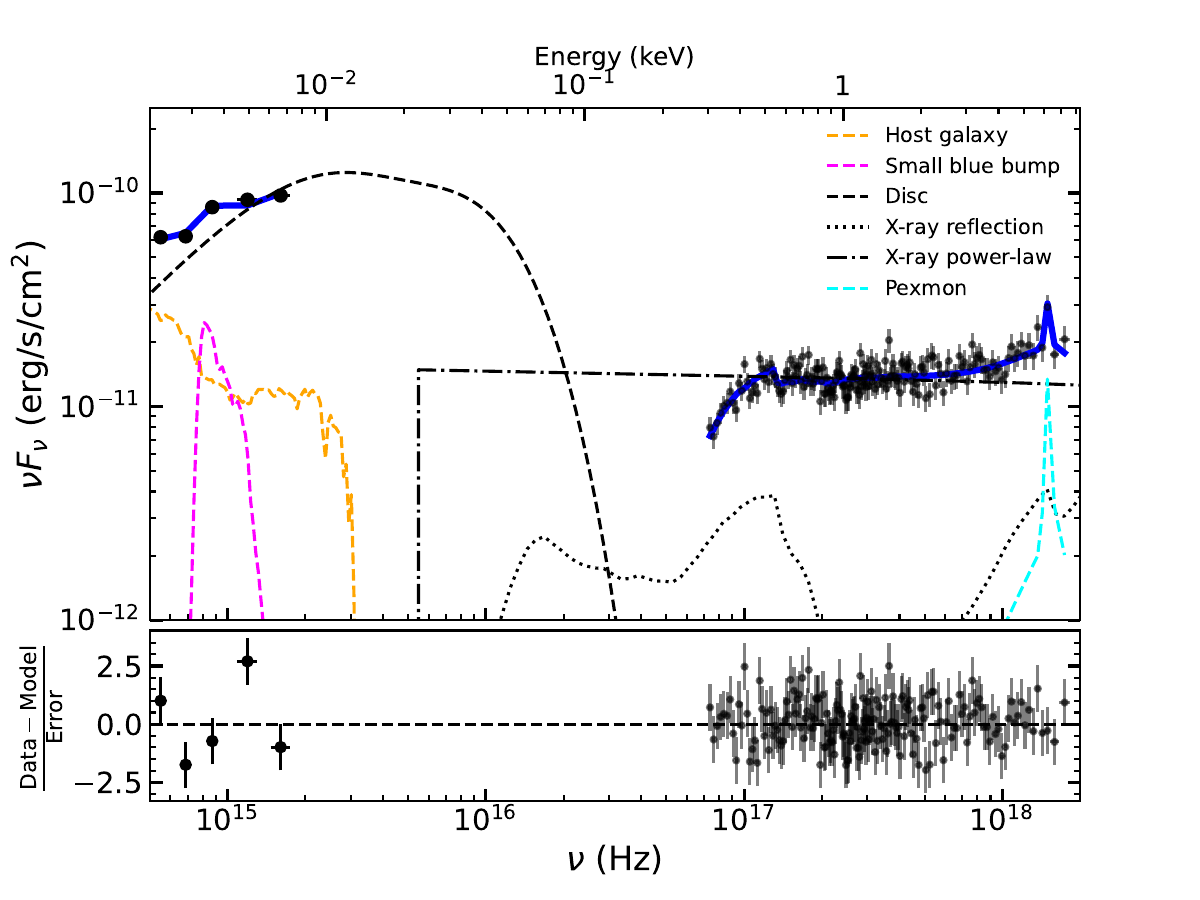}
    \caption{The average optical/UV/X-ray SED of F9. The black points show the data, and the solid blue line shows the best-fit model for \ltransf>0. We also show the different emission components (without absorption): disc (black dashed line), host galaxy (orange dashed line), Balmer/FeII template (pink dashed line), X-ray power law (black dashdot line), X-ray reflection (black dotted line), and reflection from distant, neutral material (cyan dashed line).}
    \label{fig:mean_sed}
\end{figure}

The mean broadband UV/optical/X-ray SED is shown with black circles in Fig\;\ref{fig:mean_sed}. We fit the mean SED using \texttt{XSPEC} \citep{xspec}, and the model: 
\vspace{0.2 cm}

Mean SED=\texttt{redenn $\times$ [KYNSED$_{\rm UV/opt.}$+host-galaxy + small-blue-bump]+
TBabs $\times$ [KYNSED$_{\rm X-rays}$+pexmon]}. 
\vspace{0.2 cm}

\noindent
\texttt{KYNSED} \citep{dovciak22} is a model that computes the emission from an accretion disc that is illuminated by an X-ray corona located above the central black hole (lamp-post geometry). It assumes that the corona isotropically emits in its rest frame. The energy spectrum of the continuum emission is a power law with high-energy exponential roll-off. Approximately half of the X-rays will illuminate the disc and will be either reflected or absorbed. \texttt{KYNSED} computes the thermal disc emission including the X-ray power that is absorbed by the disc, as well as the X-ray reflection component in a self-consistent way. 

Spectral component \texttt{redden},  with $E(B-V)=0.022$ \citep{Schlafly}, accounts for the 
Galactic extinction in the UV/optical bands. The corresponding photoelectric absorption in the X-rays was modelled with \texttt{TBabs} using $N_{\rm H}=2.86\times10^{20}\ \mathrm{cm^{-2}}$  \citep{hi4pi}. To account for the host-galaxy starlight in the optical bands, we included a galaxy template. Fairall 9 is classified as an SBa galaxy \citep[e.g.][]{Bentz09}, so we used the template of NGC\;3049, an SBab type galaxy, from the TRDS Brown Atlas \citep{Brown14} as the model component \texttt{host-galaxy}. We fixed the normalisation of the host galaxy template to $4.87 \times 10^{-15}$ erg/s/cm$^2$/A at a wavelength of $5100(1+z) = 5335$\AA. This is the expected host flux in the 5" aperture of the \swift\ observations (M. Bentz, private communication). This estimate was based on the detailed work on the morphology of the host galaxy of F9 \citep{Bentz09, Bentz13}. 

To fit the U-band flux well, it is necessary to include a model component for the Balmer continuum and the blended Fe II line emission, which originate from the broad line region (model component \texttt{small-blue-bump}). For this, we use the template in \citet{Mehdipour15}, with free normalisation, $N_{\rm sbb}$ (the normalisation in this case was chosen to be equal to the component flux at the centroid wavelength of the U band filter). Finally, for the X-ray spectrum, we also include a neutral reflection component from distant matter, modelled with \texttt{pexmon} \citep{nandra07}. The photon index and the high-energy cutoff of this component are tied to the respective \texttt{KYNSED} parameters. We fix the inclination to 30$^o$, the scaling factor for reflection to -1 (which corresponds to an isotropic source above the disc), and leave the normalisation ($N_{\rm pexmon}$) free to vary.

Regarding \texttt{KYNSED}, we assume a disc inclination of $\theta=30^o$, and a black hole mass of $M=2\times10^8M_\odot$. The high-energy X-ray cut-off is not well constrained for this source. \cite{Akylas21} measured $E_{\rm cut}\sim 200$ keV but with a large error, while other studies report only a low-energy limit for $E_{\rm cut}$ that is larger than $\sim 400$ keV \citep[e.g.][]{lohfink16, PanagiotouWalter20, kamraj22}. We therefore fixed $E_{\rm cut}$ at 300 keV. There are five more parameters that determine the model spectrum, which we left free to vary during the fit. These are: the black hole spin, \spin, the accretion rate, \medd, the colour correction factor, \fcol\footnote{This parameter accounts for the deviation of an accretion disc's spectrum from a pure blackbody due to electron scattering and vertical structure effects of the disc atmosphere.}, the height of the corona $h$, the photon index of the continuum spectrum $\Gamma$, and \ltransf. The latter parameter quantifies the power transferred to the corona, $L_{\rm transf}$, as a fraction of the total accretion power, $L_{\rm disc}$. If \ltransf\ is positive, then $L_{\rm transf}$ is extracted from the disc and transferred to the corona (by an unknown mechanism). If \ltransf<0, the corona is powered by an external source of power (not associated with the accretion process). 

We fitted the mean SED separately for negative \ltransf\ and positive \ltransf. Each time, we fit the data keeping \fcol\ fixed at 1, 1.7  and $-1$, in which case \fcol\ is computed according to the prescription of \citet{done12}. The total number of free model parameters in each case is seven (i.e., the 5 free \texttt{KYNSED} parameters plus $N_{\rm pexmon}$ and $N_{\rm sbb}$).

The model fits the mean SED very well both when the corona is powered by the disc and when it is powered by an external source. Figure\;\ref{fig:mean_sed} shows the best-fitting model with the solid blue line, for the case where the corona is powered by accretion (\ltransf>0). The various lines show the individual unabsorbed components. 
The model provides a similarly good fit in the case where \ltransf<0. 

The best fit results 
are listed in Table\;\ref{tab:bestfit_params}. In the case where the corona is powered by accretion (\ltransf>0), the model fits the data best for \fcol=1.7. In the case where the corona is powered by an external source, the model fits the data best for \fcol=-1. Apart from this (and \ltransf, of course) the best-fit parameters are consistent under the two assumptions for the origin of the X-ray power. We list the best-fit parameters with their 1$\sigma$ error. To estimate the error, we used the \texttt{steppar} command in \texttt{XSPEC}. We approximated the resulting $\chi^2$ versus parameter value curve with a second-order polynomial. The 1$\sigma$ confidence interval is the parameter range that results in model fits with $\Delta\chi^2=(\chi^2-\chi^2_{\rm min})< 8.1$ 
\citep{Lampton}.

For some of the model parameters, fitting the mean SED yields lower or upper limits. This is the case with the spin of the BH for \ltransf<0 (best-fit value=0.998) and the height of the corona for both \ltransf>0 (best-fit value = $4R_g$\footnote{$R_g=GM/c^2$ is the gravitational radius of a BH with mass $M$.}) and \ltransf < 0 (best-fit value = $6R_g$). 

\renewcommand{\arraystretch}{1.7}
\begin{table}[t!]
\begin{center}
\caption{Best-fit $\chi^2$ and best-fit parameters.}
\label{tab:bestfit_params}
\begin{tabular}{lcc}
\hline
Parameter & \ltransf>0 & \ltransf<0 \\
\hline
\fcol    &  1.7 & -1 \\
Spin (\spin)                 & $0.82^{+0.1}_{-0.23}$ & >0.97 \\
\medd\ & $0.11^{+0.05}_{-0.04}$ & $0.09^{+0.03}_{-0.03}$ \\
\ltransf               & $0.82^{+0.01}_{-0.05}$ & $-0.7^{+0.1}_{-0.3}$  \\
Height ($R_g$)           & $\le$11
& $\le$11 \\
 & (7.5-11) & (6.2-11) \\
$\Gamma$    & $2.02^{+0.08}_{-0.08}$ & $2.04^{+0.08}_{-0.08}$ \\
$N_{\rm pexmon}^{1}$ & $0.013^{+0.014}_{-0.013}$  & $0.014^{+0.015}_{-0.013}$ \\
$N_{\rm sbb}^{2}$ & $2.44^{+0.79}_{-0.88}$  & $2.36^{+0.82}_{-0.78}$ \\
\hline
$\chi^2$ (dof=178)       &         170   & 183 \\
\hline
\end{tabular}
\end{center}
$^{1}$ in (photons/keV/cm$^2$/s) at 1 keV, $^{2}$ in mJy
\label{table:sed_best_fits}
\end{table}
\renewcommand{\arraystretch}{1}

We note that small heights of the corona may be non-physical, since the X-ray corona has a finite radius $R_C$. Generally speaking, only configurations that satisfy $R_C + R_h < h$, where $R_h$ is the radius of the BH horizon, are physical. \texttt{KYNSED} computes $R_C$ using the fact that the photon number is conserved during 
Comptonisation \citep[see Sect. 2.2 in][]{dovciak22}. $R_C$ is not a free parameter of the model. It may be computed for the best-fit model using the \texttt{XSPEC} command \texttt{xset}. In our case, $R_c=6.5R_g$ for \ltransf>0 and $R_c=5.2R_g$ for \ltransf<0. Assuming an event horizon radius of $\sim 1R_g$ (for a black hole spin of 0.998) and requiring the corona height to satisfy the above inequality, we can put tighter constraints on the height of the corona: $\sim 7.5-11 R_g$ and $\sim 6.2-11 R_g$ for the 1$\sigma$ confidence regions for \ltransf>0 and \ltransf<0, respectively. This range of corona height values is listed in parentheses in the 6th row of Table \ref{tab:bestfit_params}.

\section{Estimation of the power spectrum}
\label{sect:psdest}

\subsection{Calculation of the Periodogram}
\label{subsec:periodogram}

We used the periodogram as the basis for the power-spectrum estimation in all bands. The periodogram, $I(\nu_j)$, was calculated using the following equation:
\begin{equation}
    I(\nu_j)=\frac{2\Delta t_{\mathrm{rf}}}{N} \left | \sum_{i=1}^{N}
    \left[F(t_{\mathrm{rf},i}) -\overline{F}\right] \cdot e^{-2\pi i \nu_j t_{\mathrm{rf},i}}\right |^2,
\label{eq:per}
\end{equation}
where $\Delta t_{\mathrm{rf}}=\Delta t_{\mathrm{int}}/(1+z)$ and $t_{\mathrm{rf},i}=t_{\mathrm{int},i}/(1+z)$, with $z$ being the redshift (i.e. all times are measured on the rest frame (rf) of the source). The periodogram is calculated at frequencies $\nu_j=j/(N\cdot\Delta t_{\mathrm{rf}})$, where $j=1,\, 2,\,\dots,\, N/2$.

For the X-ray band, we divided the points in the X-ray light curve by the mean count rate, $\overline{CR}$. Hence, the term in the sum on the right-hand side of Eq.\,(\ref{eq:per}) is $\left[ CR(t_{\mathrm{rf},i})-\overline{CR}\right]/\overline{CR}$. Consequently, the X-ray band periodogram has units of 1/(day)$^{-1}$. In the optical and UV bands,  $F(t_{\rm{rf},i})$ and $\overline{F}$ are the flux measurements and their mean (in mJy), respectively. For these bands, we did not divide the light curve points by the mean flux; therefore, the periodogram has units of mJy$^2/\mathrm{day}^{-1}$ and is representative of the amplitude of the variable component in the SED, in physical units. If we can model the UV/optical power spectra in these units, then we will know that energetically the X-ray variations can account for the amplitude of the observed UV/optical variability (in mJy). 

Another reason for not dividing the periodogram with the mean squared in the UV/optical bands is that the AGN emission in the UV and (mainly) optical bands is ``contaminated'' by the host galaxy emission. Therefore, division of the periodogram by the mean flux squared would result in a decrease of the variability amplitude with increasing wavelength, since the galaxy contribution increases at longer wavelengths. The host galaxy contribution is not an issue when computing the difference  [F(t$_i$) - $\overline{F}$] because the constant host galaxy emission contributes the same in both F(t$_i$) and in $\overline{F}$.

\subsection{Stationarity}
\label{subsec:stationarity}

When estimating the power spectrum, it is usually assumed that the underlying variability process is stationary, i.e. that its statistical properties (e.g. mean, variance, power spectrum) remain constant with time. 
The light curves show different patterns before and after MJD $\sim 58530$. This is more pronounced in the UV/optical bands. As shown in Fig.\,\ref{fig:lightcurves}, the UV/optical light curves decrease at the beginning, then the flux remains nearly constant for an extended period before showing a gradual rise. After MJD $\sim 58530$, the light curves exhibit stronger short-time-scale fluctuations. Such behaviour could indicate non-stationarity. 

To investigate the hypothesis of stationarity, we divided the interpolated light curves into two parts. These parts are indicated by the vertical dashed lines in Fig.\,\ref{fig:lightcurves}. The first part is from the start of the shaded region in Fig.\,\ref{fig:lightcurves} until MJD=58530. The duration of this light curve segment is 264 days. The second part starts with MJD 58530 and also has a duration of 264 days. We computed the periodogram of the two parts of the light curve (in X-rays and in all the other bands) and then applied the $S-$test of \cite{papadakis95} to test for stationarity. The $S-$statistic is expected to be normally distributed with mean equal to zero and variance equal to one under the null hypothesis of stationarity. However, it is important that the $S-$statistic is computed using periodogram estimates that are not affected by the Poisson noise of the two light curves, which may be different. 

For this reason, we computed the $S-$statistic using 40 periodogram estimates in the case of the X-ray and B-band light curves, i.e. up to $\sim 0.15$ day$^{-1}$ (which corresponds to a timescale of 6.5 days). The Poisson noise starts to affect the power spectra at higher frequencies, as shown by the X-ray and B-band PSDs in Fig.\,\ref{fig:HX_PSD} and \ref{fig:UVopticalpsds}, respectively. The results are $S=0.95$ and $S=2.3$ for the X-ray and B-band PSDs, respectively. In the case of the W2 PSD, the Poisson noise appears to be very low; hence we considered all frequencies when computing the $S-$statistic, which is 2.3. In the case of the W1 PSD, we found $S=2.7$ using all the periodogram points up to $0.25$ day$^{-1}$ (this corresponds to a time scale of 4 days). Finally, for the U and V band PSDs we computed the $S-$statistic up to 0.11 day$^{-1}$, since the Poisson noise is more pronounced in these cases (see the respective PSDs with the best-fit Poisson noise levels in Fig.\,\ref{fig:UVopticalpsds}). We find $S_{U}=2.6$ and $S_{V}=2$. The results are consistent with the null hypothesis within 3$\sigma$, in all cases. 

We conclude that there is no significant evidence of non-stationarity, in all bands, when we consider variations up to time scales of $\sim 9$ days (in the U and V bands), 6.5 days (in the X-ray and B-band light curves), 4 days in the W1 and 2.5 days in the W2 band light curves.
The S-statistic is systematically large (i.e. around 2-2.7) in the UV/optical bands. However, this is not surprising, given that we did not simply divide the total light curve into two, but we specifically chose to compare two parts of the light curve that visually appear to be very different. In addition, given the strong correlation between the two light curves, the $S$ values are not strictly independent. Since the $S-$ statistic is 2.3 for the UVW2 light curve, we would expect similar values for the other UV/optical light curves as well.

\subsection{The X-ray PSD}
\label{sec:xray_psd}

\begin{figure}
    \centering
    \includegraphics[width=1.1\linewidth]{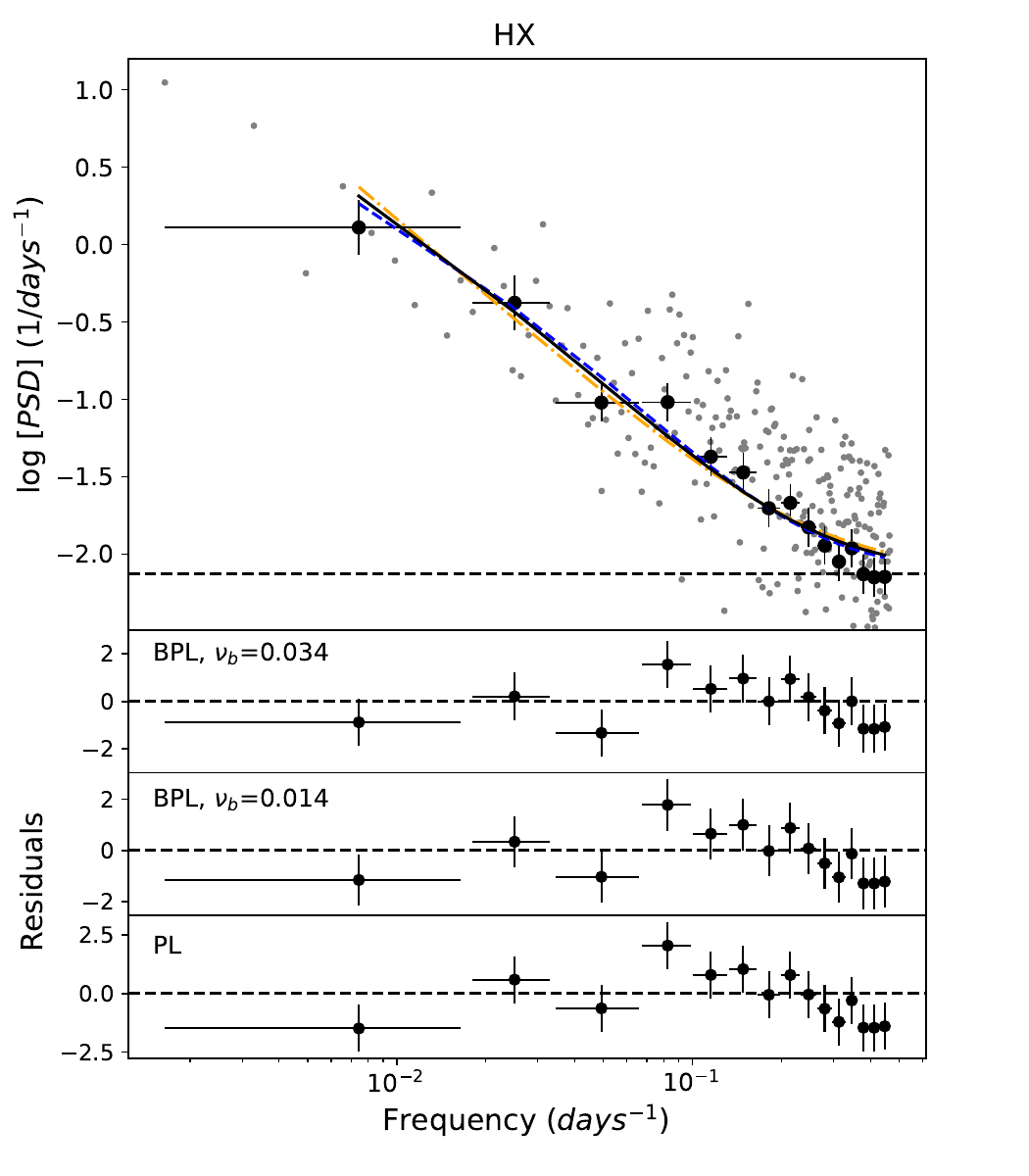}
\caption{The X-ray PSD. Top panel: Grey points show the log periodogram and black circles the binned log periodogram. The solid black and dashed blue lines show the best-fit BPL models assuming $\nu_b=0.014$ and 0.034 days$^{-1}$, respectively (see text for details). The dash-dotted orange line shows the best-fit power law model. Bottom panels: The respective best-fit residuals (i.e. (data-model)/error).}
    \label{fig:HX_PSD}
\end{figure}

The small grey circles in Fig.\,\ref{fig:HX_PSD} show the log-periodogram\footnote{When taking the logarithm of the periodogram we add an extra term of 0.25068 \citep{Vaughman} to remove the bias, i.e. $\log(PSD)=\log(I)+0.25068$, as explained in \cite{Papadakis93}} of the X-ray light curve. The scatter of the log-periodogram is large due to the fact that the periodogram is distributed as a $\chi^2$ variable with two degrees of freedom. As is customary, we binned the log-periodogram to get estimates of the PSD whose probability distribution is similar to the Gaussian distribution and their error is known and equal to $\sqrt{0.31/\mathrm{M}}$, where M the number of log-periodogram points per bin \citep[see][for details]{Papadakis93}. The mean log-periodogram is the estimate of the log PSD at the mean of the logarithm of the frequency of the individual log-periodograms, i.e. at a frequency that is the geometric mean of the frequency of the periodograms in each bin \citep[see][for details]{Papadakis93}.

The large black circles show the binned log-periodogram, calculated by dividing the log-periodogram into groups of size $M=10$ for the two lowest frequency bins and $M=20$ for all subsequent bins. We chose a smaller bin size for the lowest frequencies to extend the PSD at the lowest possible frequency. We accept the binned log-periodogram as the final estimate of the PSD. 

First, we fitted the X-ray PSD using a simple power law model of the form: log[PSD$_X(\nu)]$ = log[$A\cdot (\nu/\nu_0)^{-s}+C_{\rm PN,X}$], where $\nu_0=0.01$ days$^{-1}$. Then, we fitted the data with a bending power-law model (BPL) of the form

\begin{equation}
    \label{bending power law}
    \rm \log\left[ PSD_X(\nu) \right] = \log\left[A\cdot \nu^{-1}\cdot \frac{1}{1+(\frac{\nu}{\nu_b})^{s-1}}
    +C_{\mathrm{PN, X}}\right].
\end{equation}

\noindent
This function has a power law-like shape with a slope of $-1$ at low frequencies, which then steepens to a slope of $-s$ at high frequencies, above a bending or break frequency, $\nu_b$. 
The term $C_{\mathrm{PN, X}}$ accounts for the power level of Poisson noise. This is equal to the total variance of the Poisson noise process divided by the frequency range over which we estimate the PSD, i.e. $C_{\mathrm{PN}}=2\cdot \Delta t_{\mathrm{rf}}\cdot \overline{\sigma_{\mathrm{err}}^2}/\overline{CR}^2$, where $\overline{\sigma^2_{\rm err}}$ is the mean error squared of the light curve points. 
When fitting the X-ray PSD, we kept $C_{\rm PN,X}$ fixed at this value. 

Although the PSD plotted in Fig.\,\ref{fig:HX_PSD} is calculated over $\sim2$ orders of magnitude in frequency space, the error in the best-fit parameters is large when we leave them all free to vary during the fitting process. For that reason, we fitted the X-ray power spectrum twice, keeping the bending frequency fixed at 0.034 and $0.014\, \mathrm{days}^{-1}$, using the results of \citet{Markowitz}. These authors used RXTE light curves and calculated the 2--10 keV band power spectrum over a frequency range that is broader than ours. They determined the bending frequency with an error smaller than ours, and for that reason we decided to fix $\nu_b$ at their best-fit value (0.034 $\mathrm{days}^{-1}$) and its lower 1$\sigma$ limit ($0.014\, \mathrm{days}^{-1}$), when fitting the PSD.

\begin{table}[ht]
\caption{The BPL and PL best fit results to the X-ray PSD.}           
\label{table:Xraypsd_best_fit}     
\centering
\begin{tabular}{c c c c}        
\hline\hline                 
Model & A & s & $\chi^2$/dof \\    
      & ($1/\mathrm{days}^{-1}$) & & \\

\hline                        
BPL($\nu_b=0.034$) & 0.016 $\pm$ 0.003 & 2.10 $\pm$ 0.17 & 11.9/13 \\  
BPL($\nu_b=0.014$) & 0.024 $\pm$ 0.005 & 1.88 $\pm$ 0.13 & 14.0/13 \\
PL & 1.45 $\pm$ 0.39 & 1.64 $\pm$ 0.11 & 17.6/13 \\
\hline                                   
\end{tabular}
\end{table}

All three models fit the X-ray power spectrum well\footnote{Throughout the paper, we accept that a model provides a good fit to the data if the null hypothesis probability $p_{\rm null}$ is higher than 0.01.}. The best fit results are listed in Table \ref{table:Xraypsd_best_fit}. The solid, dashed, and dash-dotted lines in the top panel of Fig.\,\ref{fig:HX_PSD} show the best-fit models, while the bottom panels show the best-fit residuals.

\subsection{The UV/optical PSDs}
\label{subsec:uv/opticals psds}

\begin{figure*}[ht]
    \centering
    \includegraphics[width=1.1\linewidth]{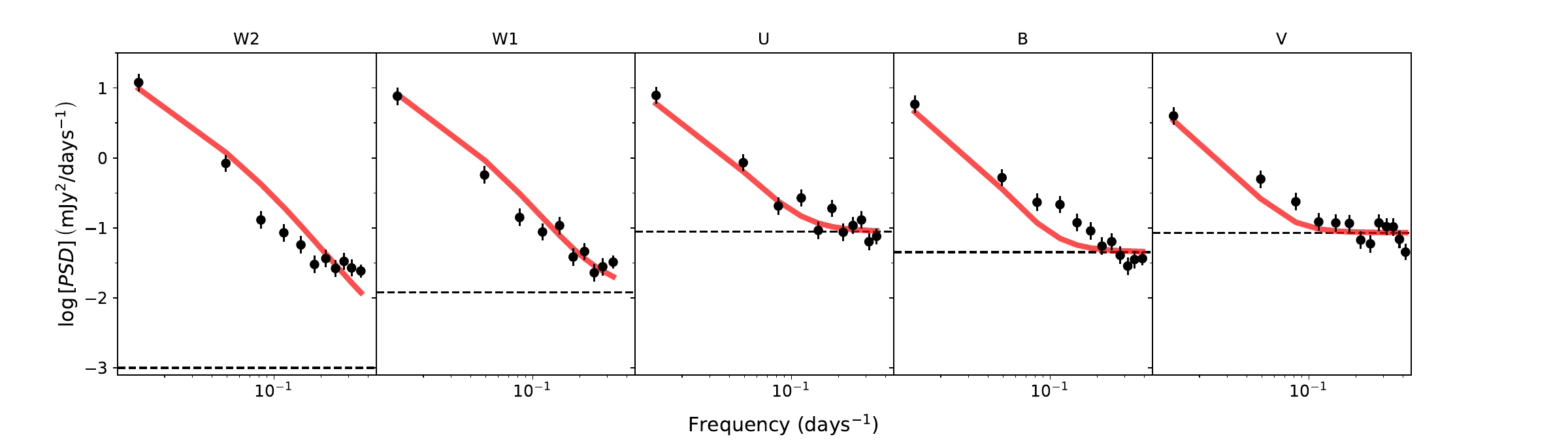}
    \includegraphics[width=1.1\linewidth]{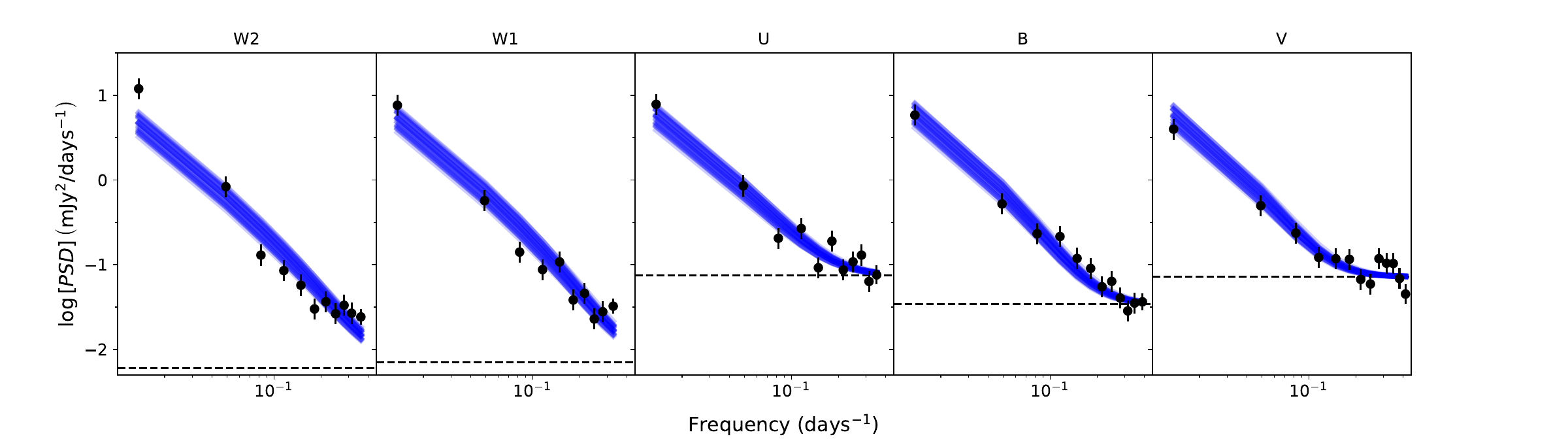}
  \caption{The solid circles show the observed PSDs in each UV/optical band. The red solid line in the top panels shows the best-fit model in the case where \ltransf>0. This model is not statistically acceptable. Blue solid lines in the bottom panels show the accepted model PSDs in the case where \ltransf<0 (see Sect. \ref{sec:fitres} for details). The horizontal dashed lines indicate the constant Poisson noise power level.}
  \label{fig:UVopticalpsds}
\end{figure*}

We estimated the UV/optical PSDs and the corresponding errors in the same way as above for the X-ray power spectrum. We calculated the periodogram using Eq.\,(\ref{eq:per}), 
and we averaged the log-periodogram estimates in bins of size 20. The solid circles in Fig.\,\ref{fig:UVopticalpsds} show the resulting binned log-periodograms, which we accept as the estimate of the intrinsic power spectrum in the UV and optical bands. These power spectra sample the observed variability of F9 on time scales from $\sim 50$ to 2.5 days. We investigate below whether the UV/optical power spectra can be explained by X-ray reverberation in the lamp post geometry, with the same model parameters that fit the mean SED. 

\section{The X-ray reverberation model power spectrum}
\label{sect: Xray_rev}

\cite{kammoun21a} studied in detail the variability in the emission of the accretion disc in the case of X-ray reverberation. \cite{Panagiotou} and \cite{Papoutsis} showed how the \cite{kammoun21a} model can be used to fit the UV/optical PSDs. We summarise below briefly the \cite{kammoun21a} model and how we can model the observed PSDs in the case of X-ray reverberation. 

We assume that the accretion disc is illuminated by the X-ray corona, which we consider to be point-like and located at a height of $h$ above the BH on its spin axis (this is the ``X-ray lamp-post'' geometry). A portion of the incident X-ray flux on the disc is reflected and re-emitted in X-rays. The rest will be absorbed and will act as an additional source of heating for the accretion disc, increasing its temperature. Assuming continuous X-ray illumination, the radiation emitted from the disc at a given wavelength and time can be expressed as a convolution of the varying X-ray luminosity with the disc response function:

\begin{equation}
    \label{eq:observed_flux}
    F_{\mathrm{em},\lambda}( t) = F_{\mathrm{NT},\lambda} + \int_{0}^{\infty} L_{X,\, \mathrm{Edd}}(t-t')\Psi_{\lambda}(t')dt',
\end{equation}

\noindent
where $F_{\mathrm{em},\lambda}(t)$ is the flux of the disc observed at wavelength $\lambda$ and time $t$. $F_{\mathrm{NT},\lambda}$ represents the steady-state flux emitted by a standard NT accretion disc, and $L_{X,\, \mathrm{Edd}}(t-t')$ is the incident X-ray luminosity, in units of $L_{\mathrm{Edd}}$. The term $\Psi_{\lambda}( t_{\mathrm{obs}})$ is the response function of the disc and quantifies the change in the emitted flux from the disc at a given wavelength $\lambda$ and observed time $t_{\mathrm{obs}}$ due to the illumination of an X-ray flash of duration $T_{\rm flash}$. It is defined as

\begin{equation}
    \label{eq:response}
    \Psi_{\lambda}(t_{\mathrm{obs}}) = \frac{F_{\mathrm{tot},\, \lambda}(t_{\mathrm{obs}})- F_{\mathrm{NT},\lambda}(t_{\rm obs})}{L_{X,\, \mathrm{Edd}} T_{\rm flash}},
\end{equation}
where $F_{\mathrm{tot},\, \lambda}(t_{\mathrm{obs}})$ is the total flux at wavelength $\lambda$ from the part of the disc that the observer sees at time $t_{\mathrm{obs}}$. This part changes with time (hence $F_{\mathrm{NT},\lambda}$, which indicates the NT disc emission from that part, also changes with time although the accretion rate remains constant). The response function is normalised to $L_{\rm X,Edd}$, the total luminosity of the X-rays in units of $L_{\mathrm{Edd}}$.


It can be shown that, if the incident X-rays are variable and Eq.\,(\ref{eq:observed_flux}) holds, then \citep[e.g.][]{Papadakis16}: 

\begin{equation}
    \label{eq:modelpsd}
   \rm  PSD_{\lambda}(\nu)=|\Gamma_{\lambda}(\nu)|^2 \cdot PSD_{X}(\nu),
\end{equation}
where $\rm PSD_{\lambda}(\nu)$ and $\rm PSD_X(\nu)$ represent the power spectrum at wavelength $\lambda$ and in X-rays, respectively. $\Gamma_{\lambda}(\nu)$ is the transfer function of the accretion disc, which is the Fourier transform of the response function:

\begin{equation}
    \label{transfer fun}
    \Gamma_{\lambda}(\nu) = \int_{-\infty}^{+\infty} \Psi_{\lambda}(t)
    \cdot e^{-2\pi i \nu t} dt.
\end{equation}


\section{Modelling the observed UV/optical power spectra}
\label{subsec:fitting_PSDs}

Equation \eqref{eq:modelpsd} provides the UV/optical power spectra at any wavelength $\lambda$ for X-ray reverberation. However, as \cite{Panagiotou} showed, it does not account for various factors that can affect the observed power spectra. First, we must take into account galactic absorption, which is particularly prominent in the ultraviolet bands. This is necessary because both the constant and variable UV/optical fluxes are reduced by absorption, so will the PSD amplitude. Secondly, the X-ray PSD plotted in Fig. \ref{fig:HX_PSD} was calculated using normalised light curves. However, $L_{\rm X,Edd}$ in Eq. (\ref{eq:observed_flux}) is the total X-ray luminosity in units of $L_{\mathrm{Edd}}$. 
It is necessary then to multiply the observed X-ray light curve normalised to the mean by the total X-ray luminosity, in units of $L_{\mathrm{Edd}}$, $L_{\rm X,Edd}$. This is equivalent to multiplying the X-ray PSD by $(L_{X, \mathrm{Edd}})^2$. 

Based on the above, the model UV/optical power spectra in the case of X-ray disc reverberation,  $\rm PSD_{m,\lambda}(\nu)$, should be given by \citep{Panagiotou}

\begin{equation}
    \label{eq:modelpsd_final}
   \rm  PSD_{m,\lambda}(\nu) = f_{\text{abs}}^2  \cdot |\Gamma'_{\lambda}(\nu)|^2 PSD_{X}(\nu) \cdot L_{\mathrm{X,Edd}}^2 + C_{PN,\lambda},
\end{equation}
\noindent
where $f_{\text{abs}}=10^{-A_\lambda/2.5}$ accounts for the absorption by the interstellar medium in our Galaxy. The extinction at each wavelength $A_\lambda$, was calculated assuming that there is no absorption in the host galaxy, using the equations in \cite{cardelli} with $R_V=3.1$ and E(B-V) = $0.022$ \citep{Schlafly}. The term $C_{\mathrm{PN,\lambda}}$ is added to account for the power level of the experimental Poisson noise, as we did before with the modelling of the X-ray PSD.

\subsection{Calculation of the disc transfer function}
\label{sec:transfer-function}

Equation (\ref{eq:modelpsd_final}) shows that we must calculate the disc transfer function and hence the disc response function as well to model the observed UV/optical PSDs. 
We used \texttt{KYNXiltr} \citep{K23rev} to compute disc response functions. The code takes into account all relativistic effects along the light path (from the corona to the disc and from the disc to the observer) and the ionisation of the disc when computing the reflected X-rays from the disc surface. The response functions of the disc were calculated at discrete times $t_k= k \cdot \Delta t_r$ for $k=1,\, 2,\,\dots, \,N_r-1$, where $N_r=8192$ and ${\Delta t}_r=0.046$ days. 
Then, we calculate the disc transfer functions, $\Gamma_{\lambda}(\nu)$, following \cite{Panagiotou}: 

\begin{equation}
    \label{discrete_transfer}
    \Gamma_{\lambda}(\nu_j) = 
    \frac{\sum_{k=1}^{N_{r}-1} \Psi(t_k)\ e^{-2\pi i \nu_j t_k}\ \Delta t_r }
    {\Delta \lambda \cdot e^{i\pi \nu_j T_{\mathrm{flash}}}\cdot \mathrm{sinc}(\pi \nu_j 
    T_{\mathrm{flash}})},
\end{equation}
 where $\nu_j = j/(N_r\cdot {\Delta t}_r)$, $j=1,2,...,N_r/2$. The transfer function is divided by the width of the corresponding band, $\Delta \lambda$. We further divide by $e^{i\pi \nu_j T_{\mathrm{flash}}} \mathrm{sinc}(\pi \nu_j T_{\mathrm{flash}})$ to account for the fact that the X-ray flash that illuminates the disc, to compute the disc response function, is not instantaneous \citep{epitropakis16}. Instead, it has a finite width duration of $T_{\mathrm{flash}}=10t_g$, with $t_g=R_g/c=GM/c^3$. 
Finally, the transfer function in Eq. (\ref{eq:modelpsd_final}) is also divided by $D_L^2$, i.e. $\Gamma'_{\lambda}(\nu)=\Gamma_{\lambda}(\nu)/D_{L}^2$, where $D_L$ is the luminosity distance of F9. This is necessary because, by default, the disc response functions are calculated for an AGN at a distance of 1\;Mpc. 

\subsection{Fitting the UV/optical PSDs}
\label{sec:fitres}

Our aim is to investigate whether the X-ray reverberation model can fit the optical/UV PSDs, in all bands, with the same model parameters that fit the mean SED as well. So,  we used Eq.\,\eqref{eq:modelpsd_final} and computed model UV/optical power spectra, $\rm PSD_{\rm mod,\lambda} (\nu)$, for each UV/optical band. The range of values of the model parameters that we considered is listed in Table \ref{table:parametersrange-forpsds}.  It is equal to the $1\sigma$ confidence region for the best-fit values, as listed in Table \ref{tab:bestfit_params}. We kept \mbh\ and the inclination fixed to the same values we assumed when fitting the mean SED. For \ltransf>0, we fixed \fcol, $\Gamma$ and the BH spin at 1.7, 2 and 0.82, respectively, as determined by the best-fit results in this case (see Sect. \ref{sect:mean_sed}). Similarly, for \ltransf<0, we fixed \fcol, $\Gamma$ and the BH spin at -1, 2.04, and 0.998, respectively. 

\begin{table}[h]
\caption{Range of model parameters for the computation of the disc response and transfer functions.}            
\label{table:parametersrange-forpsds}     
\centering
\begin{tabular}{l l l}        
\hline\hline                 
\ltransf\ & Parameter & Range of values  \\   
\hline                        
>0                   &                         &               \\
(corona powered by   &    $\dot{m}_{\rm Edd}$  &  [0.07-0.16], step=0.02  \\
the accretion disc)  &                         &  (=0.01 for the last value) \\ 
                     &    $h(r_g)$             &  [3-11], step=2\\
                     &    \ltransf\            &  [0.77-0.83], step=0.03 \\
<0                   &                         & \\  
(corona powered by   &   $\dot{m}_{\rm Edd}$  &  [0.06-0.12], step=0.02  \\      
an external source)  &                        &  \\
                     &    $h(r_g)$             &  [3-11] step=2 \\ 
                     &    \ltransf\            &  [-1,-0.6], step=0.1 \\
\hline                                   
\end{tabular}
\end{table}

For all combinations of model parameters, we computed the disc response, transfer function, and PSD models, in all bands, as explained in Sect.\, \ref{sect: Xray_rev} and \ref{sec:transfer-function}. In total, we computed 450 and 500 model PSDs for \ltransf>0 and \ltransf<0, respectively. In fact, since the X-ray PSD is well fit by three different models (see Table \ref{table:Xraypsd_best_fit}),  we computed these model PSDs three times (i,e, for each of the three best-fit models to the X-ray PSD). 

To determine whether a certain combination of model parameters, ($\dot{m}_{\rm Edd}, h,$\ltransf),  provides a good fit to the observed PSDs in all bands, we calculate a global $\chi^2$ statistic as follows:

\begin{equation}
    \label{eq:chi2}
    \rm \chi^2_{\rm total} = \sum_{\lambda} \sum_{i=1}^{N_\lambda} 
    \frac{ 
    \left( 
    \log[ {PSD}_{ \rm obs,\lambda}(\nu_i)]-\log[{PSD} _{\rm mod,\lambda}^{\rm params}(\nu_i)] 
    \right)^2
    }
    {\sigma_{\mathrm{PSD},\lambda}^2(\nu_i) + \sigma_{\mathrm{mod},\lambda}^2(\nu_i)},
\end{equation}
\noindent
where PSD$_{\rm mod,\lambda}^{\rm params}(\nu)$ is the model PSD for each set of parameter values (computed using Eq.\,\ref{eq:modelpsd_final})\footnote{ We also fitted the data by computing the model PSD at each frequency of the periodogram estimates, and then binning the logarithm of the model PSDs exactly as we did with the log-periodograms in the observed PSDs. The best-fit results (i.e. $\chi^2_{min}$ and best-fit parameters) were almost identical. This is not surprising, given the smooth, power-law-like shape of the observed PSDs.}, $N_{\lambda}$ is the number of points in the binned power spectrum in waveband $\lambda$, and $\sigma_{\mathrm{PSD},\lambda}$, $\sigma_{\mathrm{mod},\lambda}$ are the errors of the observed and model PSDs, respectively. The error in the model power spectrum, 
$\sigma_{\mathrm{mod},\lambda}(\nu)$, is due to the fact that the best-fit parameters of the PL and BPL models that fit the X-ray PSD have an error. These errors will add some uncertainty in the model UV/optical PSDs. To calculate this uncertainty, we randomly selected 500 PSD amplitude and slope values using a Gaussian distribution with mean and standard deviation equal to the best-fit values and their error as listed in Table (\ref{table:Xraypsd_best_fit}). 
The mean of the 500 resulting X-ray power spectra is the X-ray PSD which we use in Eq.\,(\ref{eq:modelpsd_final}), and the standard deviation around the mean is the error of the X-ray PSD which is used to compute $\sigma_{\mathrm{mod},\lambda}(\nu)$ in Eq.\,(\ref{eq:chi2}) with the standard error propagation formula.

The number of PSD estimates in all the UV/optical PSDs is 57. 
The model parameters that are free during the model fit are eight: $\dot{m}_{\rm Edd}, h,$\ltransf, and the 5 Poisson noise constants (one for each band). The predicted Poisson noise level based on the error of the light curve points turned out to be rather large for the optical power spectra. For example, the predicted Poisson noise for the V and B band PSDs is 0.12 and 0.06, respectively. This would imply a log-Poisson noise level of -0.92 and -1.22 mJy$^2$/day$^{-1}$, respectively. This is significantly higher than the flattening of the B and V band PSDs at high frequencies (see the two rightmost panels in Fig.\,\ref{fig:UVopticalpsds}). We suspect that this could be due to an overestimation of the light curve errors. For this reason, we left C$_{PN,\lambda}$ in Eq.\,(\ref{eq:modelpsd_final}) as a free parameter during the fitting process\footnote{ The best-fit C$_{PN,\lambda}$ values turned out to be $\sim 1.7$ and $\sim 2.8$ smaller than the predicted Poisson noise level in the B/V and in the UVW1/UVW2 PSDs, respectively. This would imply an overestimation of errors of the order of 1.3 and 1.7 in the respective bands. Interestingly, we found that the best-fit C$_{PN,\lambda}$ value in the U band PSD is almost equal to the predicted Poisson noise level.}. Consequently, when we fit all the PSDs simultaneously, the number of degrees of freedom (dof) is 49.
We consider that a set of model parameters fits well all UV/optical PSDs if $\chi^2_{\rm total}<74.9$, which corresponds to a null hypothesis probability greater than 0.01. 

We found that no combination of model parameters can fit well all UV/optical power spectra together in the case where \ltransf>0.
The lowest value of $\chi^2$ is 99.6 and is obtained for the BPL X-ray PSD with $\nu_b=0.014$ days$^{-1}$ and \medd=0.11, h=11 r$_g,$\ltransf=0.83. The solid red line in the top panel of Fig.\,\ref{fig:UVopticalpsds} shows this model power spectrum. It overpredicts the variability power in the W1 and W2 bands and underpredicts the variability power in the optical bands (B and V). We conclude that X-ray reverberation with a corona powered by the accretion process cannot explain both the mean SED and the UV/optical variability of Fairall 9, according to our model assumptions. 

On the other hand, the X-ray reverberation model for \ltransf<0 fits the data well. The solid blue lines in Fig.\,\ref{fig:UVopticalpsds} show the X-ray reverberation power spectra in the case where \ltransf<0, and the BPL model for the X-ray PSD with $\nu_b=0.014$ days$^{-1}$.  The model power spectra fit well the observed PSDs in all bands and at all frequencies. We notice that the observed power at the lowest frequency in the UVW2 PSD is higher than the model prediction, but this excess power is not statistically significant.
The range of model parameters that fit the data is: [0.06,0.12] for \medd, [7-11 r$_g$] for the corona height and [-0.6,-1] for \ltransf. This is almost the same as the 1$\sigma$ of the respective parameters from the fitting of the mean SED.


\begin{figure}[hbtp]
    \centering
    \includegraphics[width=1\linewidth]{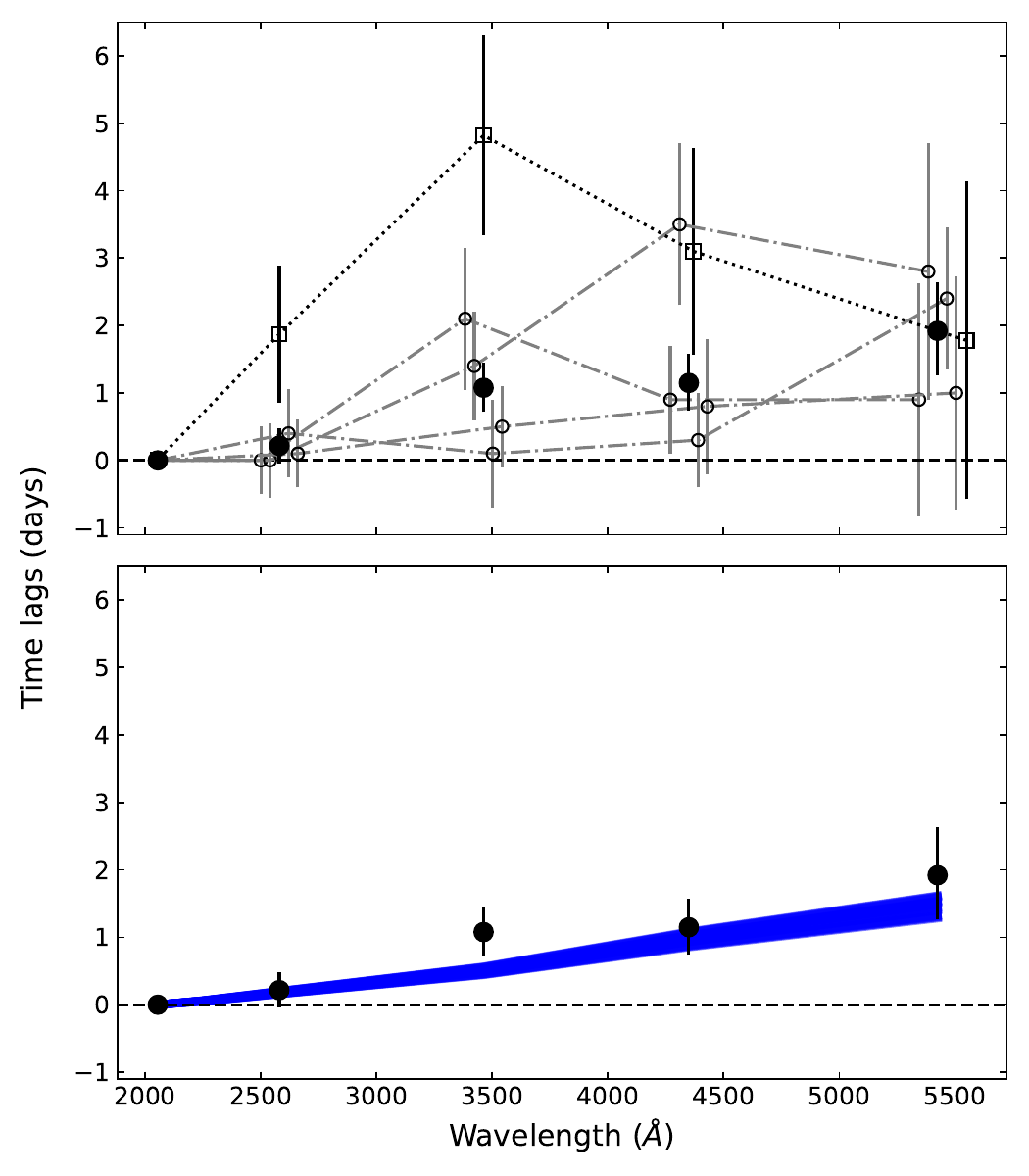}
    \caption{Upper panel: The open circles (connected with dot-dashed lines) and the open squares (connected with a dotted line) show the time-lag measurements listed in Table 5 and Table 3 of \cite{edelson}. These are time lags measured with respect to W2 from their ``quarterly analysis'' and from the full light curves, respectively. In the latter case, we show the time lags measured using the original light curves without detrending. The individual quarterly lag measurements are slightly offset in wavelength for clarity (from first to fourth quarter, left to right). The open squares corresponding to the V and B band time-lags are also slightly offset in wavelength for clarity. The errors are the ``average uncertainty on the lag'', computed as explained in the caption of Table 3 in \cite{edelson}. The filled circles show the weighted mean of the time-lags at each wavelength. Lower panel: The solid blue lines show the model time-lags in the case of X-ray reverberation for \ltransf<0, considering the model parameters that fit the SED and the UV/optical PSDs. Black circles show the weighted mean time-lag measurements as in the upper panel.}
    \label{fig:timelags}
\end{figure}

\section{Modelling the observed time lags}
\label{sect: timelags}

 According to the X-ray reverberation hypothesis,  the cross-correlation function between X-rays and the disc emission at wavelength $\lambda$, $CCF_{\lambda}(\tau)$, will be equal to

\begin{equation}
    CCF_{\lambda}(\tau) = \int_{-\infty}^{\infty} \Psi_{\lambda}(t)ACF_{X}(\tau-t)dt,
\label{eq:ccf}
\end{equation}
where $\Psi_{\lambda}(t)$ is the response of the disc at wavelength $\lambda$, and $ACF(t)$ is the auto-correlation of the X-ray variability process \citep[see e.g.][]{peterson93,welsh99}. The equation above shows that the cross-correlation between X-rays and the UV/optical light curves will be a blurred version of the response function of the disc. If the X-ray variability process was that of white noise, then $ACF_{X}$ would be a delta function and $CCF_{\lambda}$ would be identical to $\Psi_{\lambda}$. In this case, the peak of CCF$_{\lambda}$ (i.e. the ``time-lag'' between the X-rays and the disc flux at $\lambda$, $\tau_{X-rays/\lambda}$) should be representative of the centroid of the response function, 

\begin{equation}
    \label{time lags}
    \tau_{\mathrm{X-rays/\lambda}}= \frac{ \int {t\cdot \Psi(\lambda, t) dt}}{\int \Psi(\lambda, t) dt}.
\end{equation}

However, the X-ray power spectrum of nearby Seyferts shows a red noise shape, which implies that $ACF_{X}$ is broad. Consequently, $CCF_{\lambda}$ cannot be the same as $\Psi_{\lambda}$, and any characteristic peaks in $CCF_{\lambda}$ will not necessarily coincide with the centroid of $\Psi(t)$. This is one of the reasons why \cite{welsh99} suggested that the cross-correlation between the ``continuum'' (i.e. X-ray) and reprocessed emission (the UV/optical light curves in our case) should be measured after removing low-frequency variations from the light curves. 

In AGN studies it is customary to measure the cross-correlation between a light curve at a wavelength $\lambda$ and a reference band $\lambda_{ref}$, which is usually one of the shortest wavelength light curves.
Since $ACF_X$ in Eq.\,(\ref{eq:ccf}) is the same for $CCF_{\lambda}$ and $CCF_{\lambda-ref}$, we assume that even if the cross-correlation between X-rays and the light curves at $\lambda$ and $\lambda_{ref}$ is a blurred version of $\Psi_{\lambda}(t)$ and $\Psi_{\lambda-ref}(t)$, respectively, the observed time-lag between $\lambda$ and $\lambda_{ref}$, $\hat{\tau}_{\lambda,\lambda-ref}$, will be roughly equal to $\tau_{\mathrm{X-rays/\lambda}}-\tau_{\rm X-rays/\lambda-ref}$.

The open circles in the upper panel in Fig.\,\ref{fig:timelags} show the observed time-lags that \cite{edelson} measured when they divided the full light curves into 4 parts, with a duration of $\sim 162$ days each. The time-lags were measured with respect to the W2 band. They are not identical, but this is to be expected due to the red-noise nature of the UV/optical light curves. Although we have not detected non-stationarity in the light curves, this does not imply that the four sample cross-correlation functions must be identical. On the contrary, it is quite common for the sample versions of many random functions to be different in the case of a red-noise process. On the other hand, the fact that we find no indications against non-stationarity implies that the mean of the four time-lag measurements should be closer to the intrinsic time-lag between the various bands.  

The open squares in Fig.\,\ref{fig:timelags} show the sample time-lags that \cite{edelson} measured when they used the full light curves. The time-lags between the X-ray and the UV/optical bands should depend on the duration of the light curves according to Eq.\,(\ref{eq:ccf}), since $ACF_{X}$ is very broad in AGN. Consequently, the blurring of the disc response function may increase with longer light curves. However, the effect on the inter-band, UV/optical time-lags may be smaller, based on the discussion above. In fact, the full-light curve time-lags in Fig.\,\ref{fig:timelags} do not appear to be significantly larger than the quarterly lag measurements when we take into account their error. Therefore, we used the lag measurements plotted in Fig.\,\ref{fig:timelags} and computed their weighted mean, which should be more representative of the intrinsic time-lags between the UV/optical bands and W2\footnote{We note that the mean time-lags are almost the same if we consider only the four quarterly measurements.}.

We computed model time-lags using the disc response functions for the model parameters that fit both the mean SED and the UV/optical PSDs. First, we calculated the predicted time lags between the X-rays and the light curves in the UV/optical bands as follows:

\begin{equation}
    \label{discrete_timelags}
    \tau_{\mathrm{X-rays/\lambda}} = \frac{\sum_{k=1}^{N_r-1} t_k \cdot \psi(t_k,\lambda)}
    {\sum_{k=1}^{N_r-1} \psi(t_k,\lambda)}.
\end{equation}

\noindent
The time lags calculated with the equation above are in the rest frame of the source. We transform them into the observer's frame by $\tau_{\mathrm{obs}}(\lambda)=\tau_{\mathrm{X-rays/\lambda}}\cdot(1+z)$. Furthermore, since the observed time lags are measured with respect to the W2 light curve, we subtracted the model time lags for the W2 band from the W1, U, B and V time lags.

The solid blue lines in the lower panel of Fig. \eqref{fig:timelags} show the resulting model time lags. Clearly, the X-ray reverberation time-lags are fully consistent with the observations when we consider the model parameters that also fit the mean SED and the observed UV/optical power spectra. This is verified by the $\chi^2$ test: the $\chi^2$ values range between 2.2 and 4.6 with four degrees of freedom (we do not account for the time lag of the W2 band as it is set to zero). 

We notice that the observed time delay in the U band is larger than the model predictions. This difference is not statistically significant due to the large error of the observed time lag (even when we consider the weighted mean). However, this is the energy band where additional time delays may appear due to the continuum emission from the BLR. According to our best-fit results to the mean SED, $\sim 25$\% of the observed U band flux  (excluding the host galaxy contribution) may originate from the BLR continuum emission (see the pink and black dashed lines in Fig.\,\ref{fig:mean_sed}). Therefore, it is reasonable to assume that time delays due to reprocessing by the BLR will also be detected. According to our best-fit results, the continuum flux at 5100 \AA\ is 4$\times 10^{-11}$ ergs/sec/cm$^2$. This implies L$_{5100\AA}\sim 2.1\times 10^{44}$ ergs/sec. Using Eqs. (5) and (6) in \cite{Netzer22}, we predict a time delay in the U band of $\sim 0.4$ days, which is entirely consistent with the difference between the observed time-lag in the U band and the model continuum time lags shown by the blue solid lines in the lower panel of Fig.\,\ref{fig:timelags}.

\section{Summary and discussion}
\label{sect: discussion}

We studied the mean X-ray/UV/optical spectral energy distribution, the UV/optical power spectra and the time lags  of the active galaxy Fairall 9 in the context of the X-ray reverberation model. We used light curves and time-lag measurements from \cite{edelson} who presented the results of the intensive \textit{Swift} monitoring campaign of the source from mid-2018 to early-2021. This is one of the most extensive multi-wavelength AGN monitoring campaigns with \textit{Swift} so far. We calculated the mean SED and the power spectra in the UV/optical bands at frequencies $\sim 1/75$ to $1/2.5$ days$^{-1}$. Our main results are summarised as follows.

\begin{enumerate}
    \item The mean SED of Fairall 9 can be well fitted assuming X-ray illumination of the disc, both when the X-ray corona is powered by the accretion process and by an external source of power, for reasonable parameter values. 
    \item  X-ray reverberation can explain the optical/UV power spectra, assuming the same parameters that fit the SED, but only when we consider the case of the externally powered corona. 
    \item X-ray reverberation can also explain the observed time-lags, with the same model parameters that fit both the mean SED and the observed UV/optical PSDs. Although not statistically significant, we detected an excess of time delays in the U-band, which could indicate additional reverberation delays from the BLR clouds.  
\end{enumerate}

\noindent 
The main conclusion from our work is that the energy spectrum and the variability properties of Fairall 9 are consistent with a NT accretion disc, which extends to the innermost stable orbit and is illuminated by a variable X-ray source with a radius of $\sim 5R_g$, located at $\sim 7-11  R_g$ above the BH. X-ray 
illumination of the disc can fit the mean SED and X-ray reverberation (with the same parameters) can fit the UV/optical power spectra and the UV/optical time lags. This implies that the X-ray luminosity is fully capable of driving the variable part of the optical/UV SED on time scales as long as 50-100 days. The fact that the observed time lags are also explained by X-ray reverberation of the accretion disc, assuming the same physical parameters that fit the mean SED and the UV/optical power spectra, reinforces the hypothesis of X-ray reverberation in Fairall 9.

\subsection{The physical parameters of the X-ray corona and of the accretion disc in F9}

Our modelling suggests that the accretion rate in F9 should be between 0.06 and 0.12 of the Eddington limit. This is in agreement with other recent measurements of the accretion rate in F9. For example, \cite{hagen24} also estimate an accretion rate of $\sim 0.1$ of the Eddington ratio, while \cite{Gupta} fitted the broad-band SED of many unobscured Seyferts and measured a bolometric luminosity of $\sim 3.5\times 10^{45}$ erg/s for F9. This suggests an accretion rate of $\sim 0.14$ of the Eddington limit for a BH mass of $2\times 10^8$ M$_{\odot}$. 
The best-fit model to the mean SED is the one in which the colour correction factor, $f_c$, is the one prescribed by \cite{done12}. The best-fit model, which is also consistent with the power spectrum and the time-lag analysis, predicts a fast-rotating BH, with a spin larger than 0.97 (at the 1$\sigma$ level.

Our best model fits imply that the X-ray source is powered by a source that is not associated with the accretion process. The physical picture of an X-ray source located on the spin axis of the BH, which is not powered by the accretion process, is consistent with the model of \cite{ghisellini04}. These authors proposed that the central BH in radio-quiet AGN powers jets that propagate only for a short distance above the BH because the velocity of the ejected material is smaller than the escape velocity. If the ejection is intermittent, then the blobs of material will reach a maximum radial distance and then fall back, colliding with the blobs produced later and still moving outward. Dissipation of the bulk kinetic energy of the blobs by the collections could be the source of the plasma heating and, hence, for the generation of X-ray emission in these objects. According to this model, the dissipation of energy should occur along the axis of rotation of the black hole (hence justifying the assumption of the lamp post geometry) and the source of energy could be the rotational energy of the BH, and not necessarily accretion.

The model can result in fast X-ray variations (see, for example, Figure 4 in \cite{ghisellini04}). Consequently, X-ray reverberation could explain the fast UV/optical variations that we observe in the light curves of F9, without assuming intrinsic accretion rate variations in the disc on time scales much faster than the known characteristic time scales of the geometrically thin, optically thick accretion discs.

\subsection{Intrinsic accretion disc variations in F9}
\label{subsect: disc_var}

Alternative models for the explanation of the optical/UV variability in AGN involve the generation of mass accretion rate fluctuations in the accretion disc, which propagate from larger to smaller radii and can even modulate the variability of the X-rays emitted from the central region of AGN. \cite{lubarski97} was the first to propose fluctuations in the viscosity parameter at different radii in the accretion disc, which would propagate inward. These fluctuations could result in fluctuations of the accretion rate and consequently of the emitted luminosity. The  \cite{lubarski97} model was proposed to explain the optical/UV variations in AGN. According to the model, the characteristic time-scale of viscosity fluctuations is of the order of the local viscous time-scale, and the fluctuations at different radial scales are uncorrelated. 

Similar models were later proposed to explain the X-ray variability in X-ray binaries and AGN. \cite{kotov01} showed that, if the X-ray corona is accreting, then a model where accretion rate perturbations 
propagate inward on the diffusion time-scales could explain the continuum X-ray time lags in the black hole X-ray binaries. Later, \cite{areval06} proposed a similar fluctuating-accretion model to explain the spectral-timing properties of the X-ray variability of accreting black holes, including AGN.

Propagating fluctuations models have been proposed in recent years to explain the UV/optical variability of AGN. For example, \cite{mahmoud1} proposed fluctuations in the mass accretion rate in a ``warm''-Comptonisation region to explain the UV variations in Ark 120. Later, \cite{hagen24,hagen25} proposed a similar model for the explanation of the observed UV/optical variations in Fairall 9, using the same multi-wavelength data that we use in this work. In order to explain the fast UV/optical variations, \cite{hagen24}  assumed a phenomenological model, where the time scale of the fluctuations follows a power-law dependence on the disc radius (variations on the local viscous time scales, as originally proposed by \cite{lubarski97} would not work). They also assumed that the mass accretion rate fluctuations from the warm disc were suppressed by some factor before propagating into the X-ray corona. No physical mechanism was proposed for these assumptions. 

Arguably, the most important prediction of the inward propagation fluctuations is that the variations in the low-energy bands should lead the variations at higher energy bands (i.e. variations should start in the optical and then propagate to the UV bands). This is opposite to what is observed in all AGN where optical/UV time-lags can be detected. In all cases, we observe the opposite behaviour. Consequently, even in models where the disc is intrinsically variable on very short timescales,  reverberation of the continuum by some distant material (either the BLR or winds) is assumed to explain the observations. 

Interestingly, there is an indication of intrinsic disc variability in the W2 power spectrum. As we discussed in Sect. \ref{sec:fitres}, the observed power at the lowest frequency of the W2 PSD is greater than the best-fit reverberation model. Although the best-fit reverberation model for all UV/optical PSDs is statistically acceptable, this excess power in the W2 PSD is at the $\sim 3\sigma$ level. It is $\sim 0.4$ (mJy)$^2$/day$^{-1}$ (in log space), which implies that the additional variability component at this frequency contributes $\nu\times$PSD($\nu)\sim 0.03$ mJy$^2$ to the total variance, or $\sim 0.17$ mJy to the standard deviation of the light curve. The mean of the W2 light curve is 7.1 mJy, so the variability amplitude of this component is 2.4\% of the mean of the light curve. 

\begin{figure}
    \centering
    \includegraphics[width=1\linewidth]{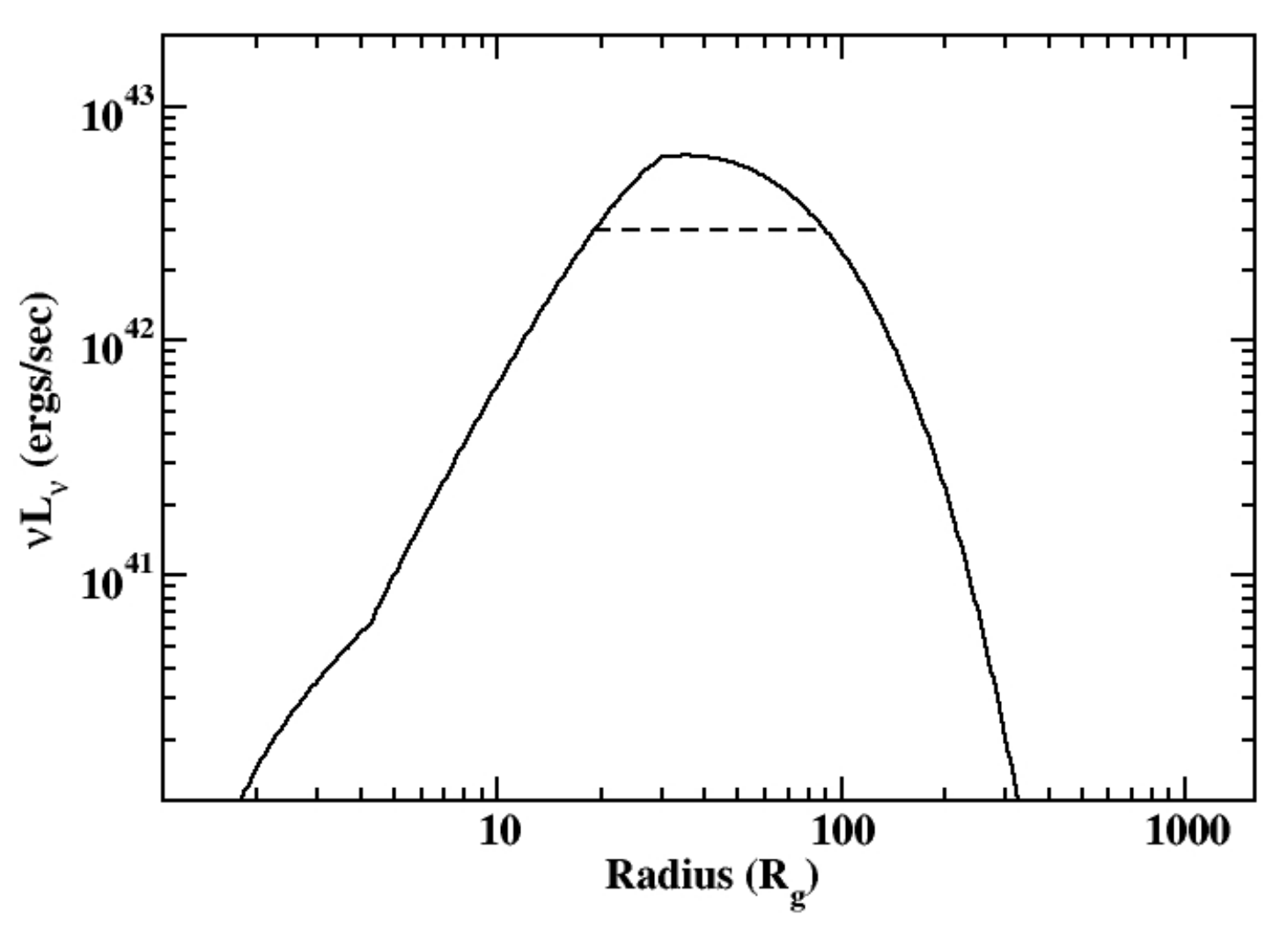}
    \caption{Flux radial profile of the disc emission in the W2 band assuming the best-fit parameters listed in Table \ref{tab:bestfit_params} for \ltransf<0.}
    \label{fig:fluxprofile}
\end{figure}

Figure \ref{fig:fluxprofile} shows the flux radial profile of the disc in the W2 band assuming the best-fit parameters for \ltransf < 0. The W2 emission does not originate from a single radius (as is usually assumed), but from a wide area of the disc. The maximum luminosity is emitted from $\sim 25-35$ $R_g$, but the horizontal dashed line shows that the disc emits more than half of the maximum luminosity from $\sim 20$ to $80$ R$_g$. The total luminosity in the W2 band is $\sim 4.9\times 10^{44}$ ergs/sec, and the total luminosity from disc emission below 15 $R_g$ is $\sim 4$\% of the W2 luminosity. 

If the additional variability that we detect in the lowest frequency of the W2 power spectrum is due to intrinsic disc variations, it could then originate from radii smaller than 15$R_g$, as long as the amplitude of the variability component would be $\sim 20$\% of the local mean. 
Intrinsic variations of the accretion rate are expected to operate on the viscous time scale, $t_{visc}$, which determines the time scale on which variations of the local density of the surface of the disc evolve. If the ratio of the disc radius to the disc height is equal to 10 and the viscosity parameter is 0.1, then, using Eq. (8) in \cite{paolillo25}, the viscous time scale in \cite{shakura73} discs is

\begin{equation}
    \label{eq: tvisc_simplified}
    t_{visc} \approx  1060 \cdot 
    \left( \frac{R}{10R_S}  \right)^{3/2}\, \mathrm{days},
\end{equation}
\noindent
where $R_S$ is the Schwarzschild radius $(R_S=2R_g)$. At radii smaller than 15$R_g$=7.5$R_S$, $t_{visc}$ is <700 days, which is long, since the extra variability appears on time scales of the order of 70 days. However, thermal ($t_{th}$) or sound crossing time scales ($t_{sound-R}$) can be up to 100 times smaller than $t_{visc}$. The first is the time it takes for the disc to reach thermal equilibrium if a thermal imbalance occurs, while the second is the timescale of the propagation of sound waves in the radial direction. Therefore, the additional low-frequency variability power we observe in the W2 PSD could be due to intrinsic variations in the innermost region of the disc which operate on the thermal or sound-crossing time scales.

We do not observe significant variability power on top of the X-ray reverberation predictions in the power spectra of the other bands, even at low frequencies. This is consistent with the hypothesis that the accretion disc varies only in its innermost region. This part of the disc will also emit radiation at longer wavelengths, but the percentage of this emission compared to the total emission in these wavelengths should decrease with increasing wavelength. On the other hand, if the disc is variable at even larger radii, the respective timescales will be much longer than the timescales we sample with the present light curves, so we would not be able to detect such an excess variability amplitude in the observed power spectra. We need longer light curves to investigate this issue. 

Finally, we note that the excess variability amplitude in the W2 band cannot be due to variations from the variable continuum emission from gas in the BLR. Such an emission should be more pronounced in the W1 and, mainly, in the U band, where we do not detect any additional variability, in addition to the best-fit X-ray reverberation model.

\begin{acknowledgements}
The authors thank the referee for helpful comments and suggestions that helped us significantly improve the paper. We thank M. Bentz for providing the measurement of the host galaxy flux within 5 arcsec at 5100 \AA. Part of the research leading to these results has received funding from the EU HORIZON-MSCA-2023-DN Project 101168906 ‘TALES: Time-domain Analysis to study the Life-cycle and Evolution of Supermassive Black Holes'. This research has used the NASA/IPAC Extragalactic Database, which is funded by the National Aeronautics and Space Administration and operated by the California Institute of Technology.

\end{acknowledgements}

\bibliographystyle{bibtex/aa}
\bibliography{refs}

\begin{appendix}
\section{PSD estimation from linearly interpolated light curves}
\label{app:simulations}

\begin{figure}[!h]
    \centering
     \includegraphics[width=1\linewidth]{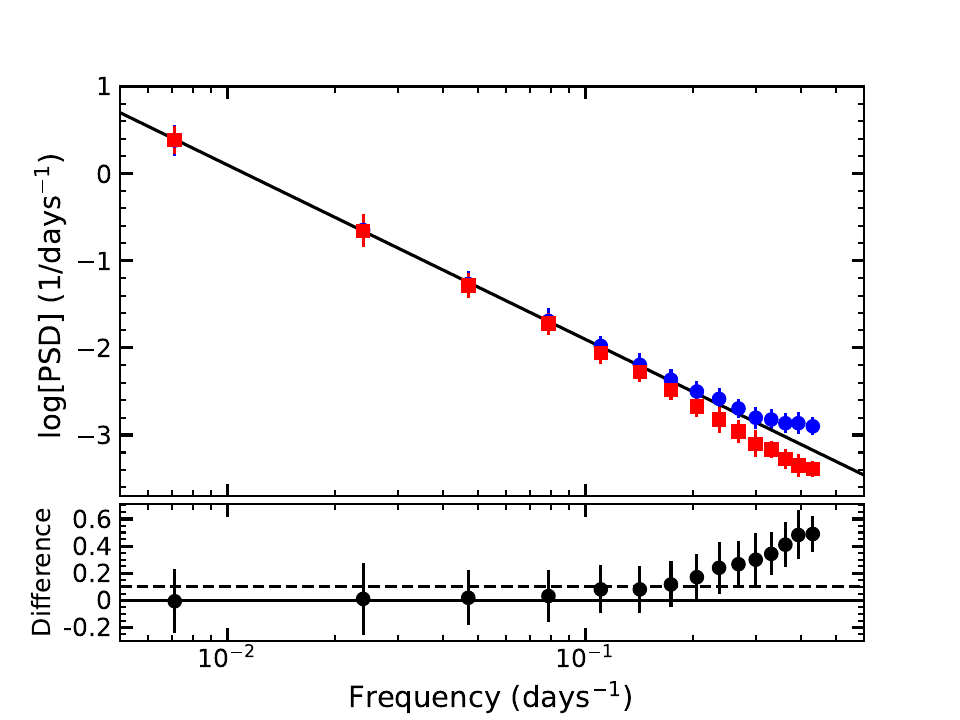}
    \caption{Power spectra computed from simulated light curves. The blue points show the power spectrum when the simulated light curves are sampled every 1.1 days, and no interpolation is performed. The red squares show the power spectrum when we use linear interpolation to create evenly sampled light curves. The black line shows the input power spectrum. 
    The black points in the bottom panel show the difference between the blue circles and the red squares in the upper panel.}
    \label{fig:simulations}
\end{figure}

In this section, we investigate whether the linear interpolation applied to the observed light curves to obtain uniformly sampled light curves introduces a bias in the estimated PSD. To this end, we generated 50 light curves using the method of \citet{TK95}, each with a duration of 637 days, matching the duration of the Fairall 9 X-ray light curve used in our PSD analysis, and a time step of 0.1 days. The simulated light curves were produced assuming an input PSD described by a power law with slopes of $-1.5,-2,-2.5$ and $-3$. In each case, the normalisation was chosen in such a way that the PSD is $1.45$ / day$^{-1}$ on $\nu=0.01$ day$^{-1}$. In this way, the amplitude of the simulated PSD at lower frequencies is similar to the amplitude of the observed X-ray PSD shown in Fig.\,\ref{fig:HX_PSD}. 

We computed the PSD of the simulated light curves using two different approaches. First, each simulated light curve was sampled at a cadence of 1.1 days (approximately the average cadence of the X-ray observations). 
We computed the periodogram and the log-periodogram of each light curve, following the procedure described in Sect.\,\ref{sec:xray_psd}. Then, at each frequency, we computed the mean and the standard deviation of the 50 log-periodograms. No interpolation is involved in this procedure since the uniformly sampled light curves are obtained by retaining every 11th point of the simulated light curves.

Second, the simulated light curves were resampled to reproduce the sampling pattern of the observed X-ray light curve, including its gaps and irregular cadence. We then applied the same linear interpolation procedure as used for the real data to reconstruct evenly sampled light curves, with a bin size of 1.1 days. We computed the periodogram and the log-periodogram of each interpolated light curve, following the analysis described in Sect.\,\ref{sec:xray_psd}, and the mean as well as the standard deviation of the 50 log-periodograms in each frequency.

The results in the case where the PSD slope is $-2$ are shown in Fig.\,\ref{fig:simulations}. The solid black line in this figure shows the input PSD. The blue circles show the logarithm of the power spectrum of the intrinsically evenly sampled light curves, log[PSD$_{\rm even}(\nu)$]. The power spectrum in this case shows aliasing effects at high frequencies, as expected. The red squares in Fig.\,\ref{fig:simulations} show the logarithm of the power spectrum of the light curves we construct using linear interpolation of the observed light curves, log[PSD$_{\rm int}(\nu)$]. The black circles in the bottom panel show the difference between the two power spectra, $\Delta$[PSD($\nu$)]=log[PSD$_{\rm even}(\nu)$]-log[PSD$_{\rm int}(\nu)$]. As expected, the power spectrum in both cases is identical to the input PSD at low frequencies. Therefore, if the observed light curve is densely sampled, with few gaps, linear interpolation does not introduce a bias in the estimation of the intrinsic PSD at low frequencies.

At frequencies higher than $\sim 0.2$ day$^{-1}$, the power spectrum of the intrinsically evenly sampled light curve (blue points) and of the linearly interpolated light curves (red points) deviate. At these frequencies,  
PSD$_{\rm int}(\nu)$ (red points) lies systematically below PSD$_{\rm even}(\nu)$ (blue points), and the difference between the two power spectra is larger than 0.1 (in log space). The first thing to notice is that PSD$_{\rm int}(\nu)$ does not show any aliasing effects. It seems that linear interpolation smooths the intrinsic light curve over the bin size of the interpolated light curve, hence the lack of aliasing effects in log[PSD$_{\rm int}(\nu)$]. In fact, the smoothing effects extend to larger bin sizes, hence the lower amplitude of log[PSD$_{\rm int}(\nu)$] compared to the input PSD at frequencies 2--3 times lower than the Nyquist frequency, $\nu_{\rm Nyq}$.
There are a few gaps longer than 2 days in the observed X-ray light curve, the longest extending to approximately 9 days. Their mean duration is roughly $\sim 5-6$ days. It appears that linear interpolation over these longer gaps results in smaller variability amplitude of the interpolated light curves at frequencies higher than $\sim$ 1/(mean gap duration) (where "mean gap duration" is the mean duration of the gaps which are longer than the bin size of the interpolated light curve).

We found the same results, i.e. $\Delta$[PSD($\nu$)]<0.1 at frequencies lower than $\sim \nu_{\rm Nyq}$ / 3 and $\Delta$[PSD($\nu)]_{max}\sim 0.4$  at $\nu_{\rm Nyq}$, irrespective of the slope of the input PSD. However, these differences are small. Fitting log[PSD$_{\rm int}(\nu)$] with a power-law model yields nearly identical best-fitting slopes and normalisations to the intrinsic PSD. Furthermore, in our case, the difference between PSD$_{\rm int}(\nu)$ and the intrinsic PSD occurs in a frequency range where Poisson noise starts to be significant. This implies that the shape of the observed power spectrum is dominated by frequencies where interpolation does not bias the estimation of the intrinsic PSD.

In summary, our results indicate that, irrespective of the slope of the power spectrum, interpolation of densely sampled light curves (where most points are separated by a quasi-constant $\overline{\Delta t}$) may result in a decrease of the PSD amplitude at frequencies higher than the mean duration of the gaps in the light curve that are longer than $\overline{\Delta t}$. In our case, this effect does not introduce bias in the estimation of the power spectrum of the X-ray light curves (i.e. in the determination of its amplitude and slope). The same holds true for the PSDs of the UV/optical light curves, which have a similar duration and sampling pattern. 

\end{appendix}

\end{document}